# Line Positions and Intensities of the $\nu_4$ Band of Methyl Iodide Using Mid-Infrared Optical Frequency Comb Fourier Transform Spectroscopy

**Ibrahim Sadiek[1*], Adrian Hjältén[1], Francisco Senna Vieira[1†], Chuang Lu[1], Michael Stuhr[2], and Aleksandra Foltynowicz[1,+]**

[1]*Department of Physics, Umeå University, 901 87 Umeå, Sweden*
[2]*Institute of Physical Chemistry, University of Kiel, 24118 Kiel, Germany*
[+]*aleksandra.foltynowicz@umu.se*

**Abstract:**

We use optical frequency comb Fourier transform spectroscopy to measure high-resolution spectra of iodomethane, $CH_3I$, in the C-H stretch region from 2800 to 3160 cm$^{-1}$. The fast-scanning Fourier transform spectrometer with auto-balanced detection is based on a difference frequency generation comb with repetition rate, $f_{rep}$, of 125 MHz. A series of spectra with sample point spacing equal to $f_{rep}$ are measured at different $f_{rep}$ settings and interleaved to yield sampling point spacing of 11 MHz. Iodomethane is introduced into a 76 m long multipass absorption cell by its vapor pressure at room temperature. The measured spectrum contains three main ro-vibrational features: the parallel vibrational overtone and combination bands centered around 2850 cm$^{-1}$, the symmetric stretch $\nu_1$ band centered at 2971 cm$^{-1}$, and the asymmetric stretch $\nu_4$ band centered at 3060 cm$^{-1}$. The spectra of the $\nu_4$ band and the nearby $\nu_3+\nu_4-\nu_3$ hot band are simulated using PGOPHER and a new assignment of these bands is presented. The resolved ro-vibrational structures are used in a least square fit together with the microwave data to provide the upper state parameters. We assign 2603 transitions to the $\nu_4$ band with standard deviation (observed – calculated) of 0.00034 cm$^{-1}$, and 831 transitions to the $\nu_3+\nu_4-\nu_3$ hot band with standard deviation of 0.00084 cm$^{-1}$. For comparison, in the earlier work using standard FT-IR with 162 MHz resolution [Anttila, et al., J. Mol. Spectrosc. 119 (1986) 190-200] 1830 transition were assigned to the $\nu_4$ band, and 380 transitions to the $\nu_3+\nu_4-\nu_3$ hot band, with standard deviation of 0.00085 cm$^{-1}$ and 0.0013 cm$^{-1}$, respectively. The hyperfine splittings due to the $^{127}I$ nuclear quadrupole moment are observed for transitions with J≤2×K. Finally, intensities of 157 isolated transitions in the $\nu_4$ band are reported for the first time using the Voigt line shape as a model in multispectral fitting.

---

\* Currently at Leibniz Institute for Plasma Science and Technology, 17489 Greifswald, Germany
† Currently at VTT Technical Research Centre of Finland Ltd, Tekniikantie 1, 02150 Espoo, Finland

## 1. Introduction

Atmospheric monitoring of photochemically active substances such as halogenated volatile organic compounds (HVOCs) is essential for modelling the natural cycling of climate relevant trace gases [1]. Iodomethane ($CH_3I$) – a naturally occurring HVOC – is classified by the World Metrology Organization (WMO) as a very short-lived substance (VSLS), i.e., a trace gas with a local lifetime comparable to or shorter than the tropospheric transport time scale [2], and hence its distribution is non-uniform in the troposphere. Despite the fact that naturally produced $CH_3I$ has a very low atmospheric mixing ratio of 0.1−2 parts-per-trillion (ppt) [3], it is an important carrier of iodine from the ocean to the atmosphere and plays a crucial role in the chemistry of the atmosphere. When it is photodissociated by UV light, it can contribute to aerosol formation in the troposphere [4]. Despite its associated risk of exposure to workers and residents, iodomethane is still used in several industrial and agricultural applications, e.g., as a synthetic agent in organic synthesis [5] or a fumigant in buildings and soils [6, 7]. The use of iodomethane as an agricultural pesticide was promoted following the control of methyl bromide for such purposes under the Montreal Protocol in 1989 [8]. In nuclear power plants, the possible emission of radioactive iodine in the form of methyl iodide during a core melt accident is a major concern [9]. Relevant to all of these applications is the capability to install suitable leak detectors as well as to monitor personal exposure limits, where current occupational safety guidelines set the personal exposure limit to 0.3−5 parts-per-million (ppm), depending on the regulatory authority [10].

Currently, the detection of $CH_3I$ is based on gas chromatography coupled with mass spectrometry or electron capture detection (GC−MS/ECD). These techniques allow only for discrete measurements, require calibration, and are time consuming (i.e., several tens of minutes per sample). Thus, they are not suitable for workspace monitoring, leak detection, or process studies that examine rapid changes associated with natural production and loss processes. Laser-based absorption techniques should, in principle, be able to overcome the limitation of the GC-MS/ECD techniques as they provide non-invasive sensing with fast acquisition and high-sensitivity. So far, there is only one laser based absorption study[11], using mid-infrared continuous wave cavity ringdown spectroscopy (mid-IR cw-CRDS), where $CH_3I$ from a tank-purging experiment was detected at the ${}^rR_2(15)$ ro-vibrational absorption transition. A limit of detection of 15 part-per-billion (ppb) was achieved, sufficient for practical applications where ppm mixing ratios are expected, such as workplace monitoring. A major challenge that limits the applicability of laser-based absorption techniques, particularly those based on narrowband cw lasers, is the potential interference of other absorbing molecules (i.e., the so-called absorption cross-sensitivity). Therefore, broadband high-resolution measurements of mid-IR spectra of $CH_3I$ are a prerequisite for the selection of a suitable detection range, and for accurate determination of band parameters and spectral line parameters.

There exist several microwave [12, 13] and Doppler-free double-resonance [14, 15] spectroscopic studies of the ground and first excited states of $CH_3I$. In the infrared region, the molecule has been a subject of numerous studies concerning mostly the line positions of its vibrational bands: $\nu_1$ [16], $\nu_2$ [17], $\nu_3$ [18], $\nu_4$ [19, 20], $\nu_5$ [21], and $\nu_6$ [22]. Line positions of several overtone and combination bands were also investigated [22, 23]. In contrast, there exist only few studies of the line intensities of these bands of $CH_3I$. To the best of our knowledge, high-resolution studies of the line intensities exist only for the $\nu_6$ and the $2\nu_3$ bands [24, 25] and for some lines of the $\nu_5$ and $\nu_3+\nu_6$ bands [26]. There are low-resolution measurements of the overall band intensities of the six fundamental vibrations [27, 28], and *ab initio* computed intensities of its fundamental bands [29].

The $\nu_4$ band, centered at 3060 cm$^{-1}$, is interesting for spectroscopic detection as it lies in the range where many laser sources became recently available, but it is rather understudied. The two previous studies [19, 20] used standard Fourier transform infrared spectroscopy (FT-IR) with a resolution of ~0.0054 cm$^{-1}$ (or ~160 MHz), which is larger than the Doppler width of $CH_3I$ at 3000 cm$^{-1}$ and 296 K (96 MHz), and provided line positions but no information about the line intensities. In the first measurements of the $\nu_4$ band by Connes et al. [19] over 500 lines were fitted with a standard deviation (obs. – calc.) of the assigned lines of 0.005 cm$^{-1}$. However, only limited parts of the band were used and striking perturbations, causing irregularities in the sub-bands ${}^PQ_6$ and ${}^RQ_5 - {}^RQ_7$ remained unexplained. Later on, Anttila et al. [20] reanalyzed the $\nu_4$ band and were able to explain the observed irregularities in rotational structures of the sub-bands as a consequence of Coriolis and Fermi resonances with combination band levels. Overall, 1850 transitions were assigned by Anttila et al. [20] for the $\nu_4$ band and 380 transitions for the nearby $\nu_3+\nu_4$- $\nu_3$ hot band, with an average standard deviation of 0.00083 cm$^{-1}$ and 0.0013 cm$^{-1}$, respectively.

The scope of this work is to use mid-IR comb-based Fourier transform spectroscopy (FTS) to measure high-resolution broadband spectra of iodomethane ($^{12}CH_3I$) in the C-H stretch region from 2800 to 3160 $cm^{-1}$. The comb-based FTS allows acquisition times orders of magnitude shorter than conventional FT-IR spectroscopy based on incoherent light sources [30]. In addition, by precisely matching the nominal resolution of the FTS to the comb repetition rate [31-33], the comb lines are sampled accurately and high-resolution molecular spectra are measured with no observable instrumental line shape distortions, even when the absorption line widths are narrower than the nominal resolution of the spectrometer. The broadband and the high resolution capability of comb-based FTS allowed us to create a new list of the $v_4$ band and the nearby $v_3+v_4$-$v_3$ hot band based on available microwave data and the earlier model of Anttila et al. [20], and to observe the hyperfine splitting due to the large $^{127}I$ nuclear quadrupole moment for the first time in the $v_4$ band. Overall, 2603 transitions are assigned for the $v_4$ band with standard deviation of 0.00034 $cm^{-1}$ and 831 transitions are assigned for the $v_3+v_4$-$v_3$ hot band with a standard deviation of 0.00084 $cm^{-1}$. Moreover, we report the so far lacking intensities of a first set of 157 well-resolved lines in the $v_4$ band using multispectral fitting with the Voigt lineshape function.

## 2. Experimental setup and procedures

Figure 1 depicts a schematic of the experimental setup, which consisted of a mid-IR frequency comb, a multipass absorption cell, a Fourier transform spectrometer, and a gas supply system. The mid-IR frequency comb was produced via difference frequency generation (DFG) in a Mg-doped periodically poled lithium niobate crystal (MgO:PPLN) between the output of a high-power 125 MHz Yb-doped fiber laser (Menlo, Orange) and a Raman-shifted soliton generated from the same source in a highly nonlinear fiber (HNLF). The center frequency of the idler was tuned to 2975 $cm^{-1}$, where the comb has a bandwidth of 360 $cm^{-1}$ and output power of ~120 mW. The design and characteristics of this source were presented in detail by Sobon et al. [34]. Three changes were implemented to the DFG source compared to the first demonstration. First, to increase the long term stability of the idler power and center frequency, the single-mode HNLF has been replaced by a polarization-maintaining HNLF that allowed achieving the same center wavelength of the Raman-shifted soliton at a lower input power. Second, to reduce the intensity noise of the idler, the delay stage in the pump arm was stabilized using an approach similar to the one reported by Silva de Oliveira et al. [35], as shown in the lower part of Fig. 1. A fraction of the idler beam was sampled by a pellicle beamsplitter, dispersed by a diffraction grating, and a selected wavelength region was imaged on a photodetector. The idler intensity noise in the 10-300 kHz bandwidth was measured using a demodulating logarithmic amplifier (DLA). A 7 kHz dither was applied to the pump diode current of the Yb-doped fiber oscillator, and the output of the DLA was demodulated at this frequency using a lock-in amplifier to produce an error signal. A proportional-integral (PI) controller provided correction signal via a high-voltage amplifier (HVA) to the piezoelectric transducer (PZT) on which the retroreflector in the pump arm was mounted. The correction signal was monitored using a computer and kept within the operating range of the PZT by moving the translation stage on which the retroreflector assembly was mounted. Third, the repetition rate of the comb was stabilized to a tunable direct digital synthesizer (DDS) referenced to a GPS-disciplined Rb oscillator (with a relative stability of $10^{-11}$ @ 1 s), as shown in the upper left part of Fig. 1. The RF signal at the $f_{rep}$ was first mixed with a reference 100 MHz signal from DDS1, and the resulting signal was band-pass filtered at 25 MHz and further mixed with a 25 MHz signal from DDS2. The resulting error signal was fed into a PI controller that acted on the intra-cavity PZT of the Yb-doped fiber oscillator to stabilize the repetition rate. The stepping of $f_{rep}$ during spectral acquisition was done using DDS2. We note that the carrier envelope offset frequency of the DFG source is zero, because both the pump and the signal combs come from the same source.

The idler beam was led to the absorption cell via a single-mode $ZrF_4$ fiber. The absorption cell was a Herriot type astigmatic cell with 76 m path length (Aerodyne, AMAC-76LW). Analytical grade methyl iodide (Sigma-Aldrich, 99%, containing copper as stabilizer) was used as supplied without further purifications. For all the measurements the sample was contained in an Ace glass tube and was brought into the evacuated gas cells by its vapor pressure, as shown in the upper right part of Fig. 1. The headspace of the sample container was equilibrated and evacuated several times in order to have pure $CH_3I$ gas for the measurements. Dry air was used as a buffer gas. The cell pressure and the gas mixing ratio were controlled by a needle valve and a calibrated pressure transducer (CERAVAC, CTR 101 N) with a specified relative accuracy of 0.15% and readout resolution of 10 µbar. All spectra were recorded at room temperature of (296 ± 0.6) K. The ambient laboratory temperature was monitored by a PT100 platinum resistance thermometer placed in close contact with the measurement cell.

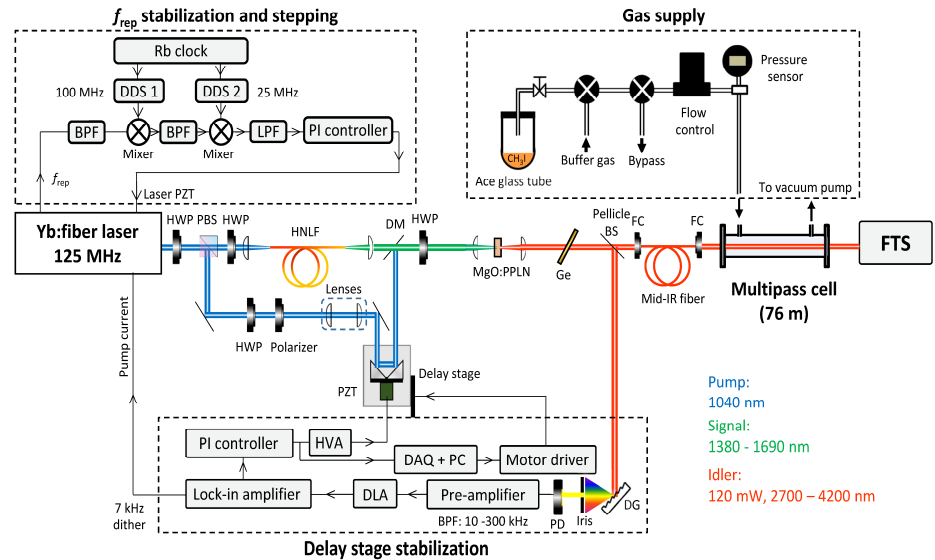

Fig.1 Schematic of the experimental setup. DDS: direct digital synthesizer, BPF: band pass filter, LPF: low pass filter, PZT: piezoelectric transducer, HWP: half-wave plate, PBS: polarizing beam splitter, HNLF: highly nonlinear fiber, DM: dichroic mirror, MgO:PPLN: Mg-doped periodically poled lithium niobate, Ge: Germanium filter, FC: fiber collimator, FTS: Fourier transform spectrometer, DG: diffraction grating, PD: photodetector, DLA: demodulating logarithmic amplifier, HVA: high voltage amplifier, DAQ+PC: data acquisition card and a personal computer.

After passing through the absorption cell, the comb beam was coupled into a home-built fast-scanning FTS [36] with scan speed of 0.4 m s$^{-1}$ and a maximum optical path difference (OPD) of 2.8 m, corresponding to a nominal resolution of 0.0037 cm$^{-1}$, or 110 MHz. The OPD was calibrated using a stable cw reference diode laser with wavelength $\lambda_{ref}$ = 1563 nm. Two out-of-phase comb interferograms were detected at the two outputs of the FTS using an auto-balanced HgCdTe detector (VIGO System, PVI-4TE-6) to reduce the intensity noise [37]. The comb and cw laser interferograms were acquired using a digital oscilloscope (National Instruments, PCI-5922) and home-written LabVIEW™ program. Afterwards, the comb interferogram was resampled at the zero-crossings and extrema of the cw laser interferogram using a home-written MATLAB® program and saved for further analysis.

The comb interferograms were analyzed using the sub-nominal resolution approach [31, 33] to achieve comb mode width limited resolution. A total of 200 interferograms with nominal resolution matched to the comb $f_{rep}$ (i.e., ~125 MHz) were acquired for a given $f_{rep}$. Next, the $f_{rep}$ was tuned by 15 Hz and the measurement was repeated. The final spectra were obtained by interleaving 13 individual spectra (each averaged 200 times, 8.3 min acquisition time) recorded with different repetition rates, yielding a sampling point spacing of 11 MHz in the optical domain. The spectra of CH$_3$I were collected as follows: First, a background spectrum was measured with the evacuated cell for one $f_{rep}$ value. Then, CH$_3$I was introduced into the absorption cell (either pure or diluted by air) and spectra were recorded for the consecutive $f_{rep}$ steps, each averaged over 200 scans. Each of the steps was normalized to the common background to yield a transmission spectrum (note that a single background spectrum could be used for all steps since the background features were broad enough to be practically unaffected by the shift in sampling point frequencies), and the absorption coefficient was calculated using the Lambert-Beer law. To eliminate the residual baseline, the absorption lines were masked and a sum of a fifth order polynomial and a series of sine terms was fit to the remaining spectral structure. The sine frequencies were chosen to match the etalon fringes that were not cancelled by the normalization to the background spectrum. Subtraction of the baseline yielded the final absorption spectrum. Finally, the spectra from the consecutive $f_{rep}$ steps were interleaved.

As explained by Rutkowski et al. [33], the sub-nominal resolution approach relies on precise matching of the FTS sampling point spacing, $f_{FTS}$, to the repetition rate, $f_{rep}$. While $f_{rep}$ can be stabilized with high accuracy (here 10$^{-11}$ relative accuracy), the accuracy of $f_{FTS}$ is given by the accuracy with which the cw reference laser wavelength, $\lambda_{ref}$, is known, which, in turn, is limited by the FTS alignment, beam divergence, fluctuations of the refractive index of the air filling the FTS etc. Any mismatch between $f_{FTS}$ and $f_{rep}$ manifests itself in the measured spectra as a characteristic instrumental line shape (ILS) distortion with odd symmetry and amplitude proportional to that mismatch [33]. Following the procedure described by Rutkowski et al. [33], we iteratively changed the value of $\lambda_{ref}$ until ILS became invisible. This was done by fitting a Voigt model to selected isolated lines across the spectrum measured at 0.11 mbar of pure $CH_3I$ and observing the root mean square (RMS) of the fit residual. The optimum value of $\lambda_{ref}$ corresponds to the minimum residual RMS. From the noise level on the baseline of this spectrum we estimate the remaining relative uncertainty in $\lambda_{ref}$ to be $4 \times 10^{-8}$, which translated to 1.5 MHz uncertainty in the $CH_3I$ line positions. This uncertainty is well below the standard deviation of 10 MHz of the assigned transitions (vide infra).

## 3. Results and discussion

### 3.1. High-resolution spectra measurements

Figure 2(a) presents the high-resolution spectrum of pure $CH_3I$ measured at 0.03 mbar in the region from 2800 $cm^{-1}$ to 3160 $cm^{-1}$. In this spectral window, $CH_3I$ exhibits a complex ro-vibrational spectrum with three distinct vibrational features, which are all measured simultaneously by our comb-based FTS spectrometer. Panels (b), (c), and (d) show zoomed in parts of the three features, revealing their dense but well-resolved ro-vibrational structure. Note that the spectra in panels (b) and (d) were measured at 0.11 mbar $CH_3I$, while the spectrum in panel (c) was measured at 0.03 mbar $CH_3I$ [same as in panel (a)] to compensate for the different intensities of the bands. The vibrational features extending from 2800 to 2900 $cm^{-1}$ [zoom in in panel (b)] were investigated by Lattanzi et al. [23] and assigned to several overlapping parallel vibrational bands including the $2\nu_5$, $\nu_3 + \nu_5 + \nu_6$, and the $2\nu_3 + 2\nu_6$ bands. The strongest band in this region [zoom in in panel (c)] corresponds to the fundamental $\nu_1$ band of the symmetric C-H stretch ($CH_3$ s-stretch), centered at 2971 $cm^{-1}$ (note that the strongest lines of the Q branch are saturated). This band was measured and assigned first at a resolution of 0.04 $cm^{-1}$ and later at 0.0054 $cm^{-1}$ by Paso et al. [16, 38]. Finally, the fundamental $\nu_4$ band [zoom in in panel (d)] corresponds to the degenerate asymmetric C-H stretch ($CH_3$ d-stretch). This band was first measured by Connes et al. [19] and further investigated by Anttila et al. [20] with a resolution of 0.0054 $cm^{-1}$. There is no information available about the line intensities or pressure broadening coefficients from previous spectroscopic studies on the three vibrational features in this region. Because of its well-resolved structure, promising for future monitoring applications, the $\nu_4$ band is selected here for further analysis to determine line positions and line intensities. As shown in Figure 2(a), the lower wavenumber tail of the $\nu_4$ band overlaps with the stronger $\nu_1$ band. Therefore, the range from the 3010 to 3160 $cm^{-1}$, which includes the Q sub-branches from $^PQ_6$ to $^RQ_{11}$, is used for determination of the band parameters in our study.

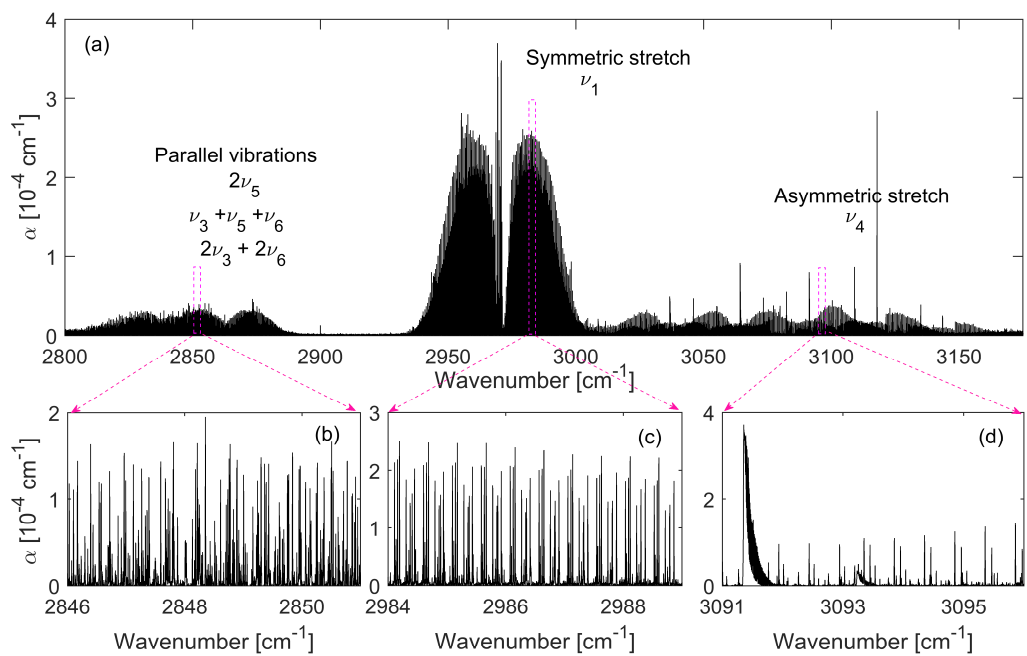

Fig. 2 (a) Overview of the broadband high-resolution spectrum of pure $CH_3I$ measured at 0.03 mbar in the range from 2800-3160 $cm^{-1}$. (b)-(d) The zoom in spectra showing the dense ro-vibrational structures of the different bands. Spectra in panel (b) and (d) were measured at a pressure of 0.11 mbar to increase the absorption signal due to the lower intensity compared to the $\nu_1$ band in panel (c).

## 3.2. Spectral simulations and assignment

Iodomethane was simulated as a symmetric top molecule with an equilibrium structure of the $C_{3v}$ point group (see Fig. 3). Three of the 3N-6 vibrational modes (i.e., $\nu_1$, $\nu_2$, $\nu_3$) are nondegenerate with A1 symmetry, while the remaining $\nu_4$, $\nu_5$, and $\nu_6$ are degenerate with E symmetry.

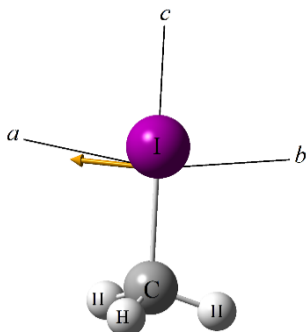

Fig. 3 The molecular structure of $CH_3I$ with the principal axes system. The yellow arrow represents the perpendicular dipole moment change to the principle c-axis as a result of the asymmetric C-H stretch vibration, the $\nu_4$ band.

The spectra were simulated and assigned using PGOPHER [39]. Figure 4(a) shows an overview of the measured absorption coefficient of 0.11 mbar of pure $CH_3I$ (black) together with the simulations of the $\nu_4$ band (blue) and the $\nu_3+ \nu_4$- $\nu_3$ hot band (red) obtained from PGOPHER. For the simulations, the input parameters of the ground states were taken from Refs. [40, 41], while for the upper states they were taken from earlier simulations by Anttila et al. [20]. Starting from the initial simulated spectra, we first assigned the K(J)=0 lines near the band origin, then we extended the assignment up to $K$=12 and $J$=75 quantum numbers. As shown in Fig. 4, an overall very good match between the experiment and the simulations is obtained. The $\nu_4$ band shows a typical structure of a perpendicular band with $Q_k(J)$ sub-bands degrading towards higher frequencies (i.e., blue shaded), indicating that the term $B' - B''$ has a small positive value. Panel (b) is an enlarged portion of the spectrum around the $Q_1(J)$ sub-band of the $\nu_4$ band. A series of weak rotational features observed between the main Q-clusters are attributed to the $\nu_3+\nu_4-\nu_3$ hot band, as previously reported by Anttila et al. [20], with intensity equal to ~7.6 % of the fundamental $\nu_4$ band, as predicted by the Boltzmann distribution factor.

Because of the large value of the $^{127}I$ nuclear quadrupole moment (with nuclear spin of 5/2), clusters of hyperfine subcomponents are expected to arise. Such hyperfine structures were previously observed in the microwave region and in the infrared region at 11 μm [14, 22, 42]. Figure 5 shows a zoom in of the measured spectrum around 3074.9 $cm^{-1}$ where clear signs of hyperfine splitting can be observed in the $^rR1(2)$ transition (highlighted in the figure). The experimental data is presented together with simulations without accounting for the hyperfine splitting [panel (a)] and after accounting for it [panel (b)]. To account for the hyperfine splitting, we simulated the spectra as described earlier in this section and included the hyperfine constants of the $^{127}I$ nuclear quadrupole moment. The nuclear quadrupole constants were fixed to those of the $\nu_6$ band taken from Carocci et al. [14], since the corresponding values for the $\nu_4$ band have not been reported in the literature. As shown in panel (b), the hyperfine subcomponents are clearly visible for the $^rR1(2)$ transition, for which $J \le 2×K$, with splitting of 6.3 × $10^{-3}$ $cm^{-1}$, while for transitions with $J > 2×K$ the hyperfine subcomponents show much smaller splitting For example, hyperfine subcomponents in the $^rR0(20)$ and $^rP2(15)$ transitions are separated by 8.35 × $10^{-5}$ $cm^{-1}$ and 1.15 × $10^{-3}$ $cm^{-1}$, respectively. Due to the small splitting of the hyperfine subcomponents for transitions with $J > 2×K$ (which represents the majority of the strong ro-vibrational transitions of the $\nu_4$ band) and the poorly resolved hyperfine splitting of transitions involving $J \le 2×K$, we have excluded hyperfine structures from spectral assignment. Only transitions with $J > 2×K$ are included in the fitting pool, while transitions showing any sign of splitting were excluded from the assignment. On the other hand, we explicitly accounted for the Coriolis and Fermi resonances, which are rather strong and perturb the $\nu_4$ band, as described by Paso et al. [16]. It should be noted that the unresolved hyperfine splitting may results in a slight asymmetry in absorption profiles, and hence affect the accuracy of the reported line intensities and positions.

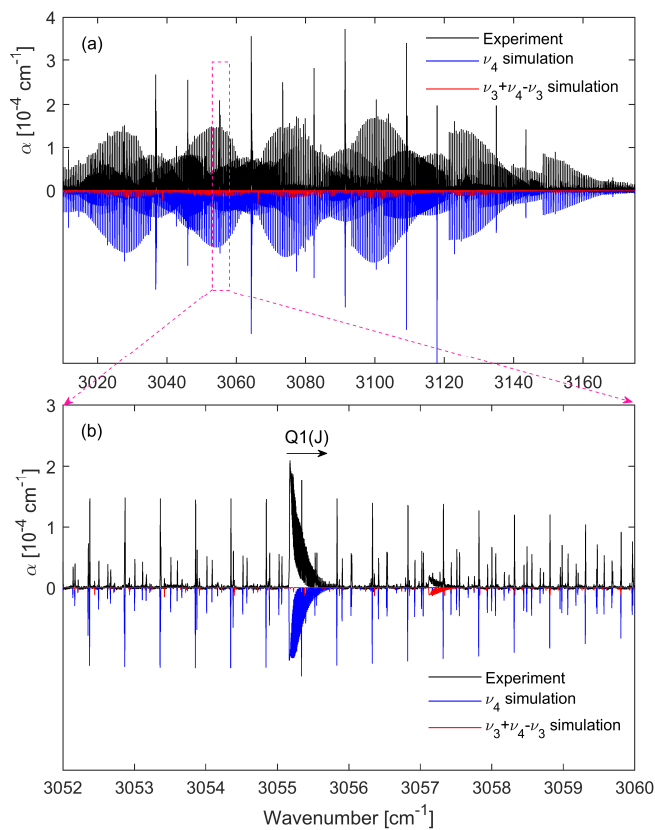

Fig. 4 (a) Measured absorption coefficient, $\alpha$, of 0.11 mbar of pure $CH_3I$ (black) at 296 K and simulations of the $\nu_4$ band (blue) and the $\nu_3+\nu_4-\nu_3$ hot band (red) obtained from PGOPHER. (b) Zoom in around the $Q_1(J)$ sub-branch of the $\nu_4$ band. The arrow indicates the degradation direction with increasing J.

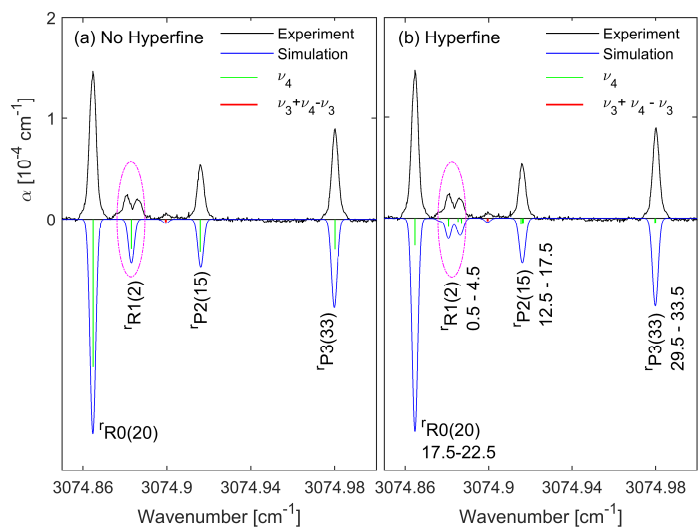

Fig. 5 Measured absorption coefficient, $\alpha$, of 0.11 mbar of pure $CH_3I$ (black) at 296 K together with line positions of the $\nu_4$ band (green) and the $\nu_3+\nu_4-\nu_3$ hot band (red) without accounting for hyperfine splitting [panel (a)] and after accounting for the hyperfine splitting [panel (b)]. The overall simulation obtained from PGOPHER is shown in blue. The ellipse highlights the $^rR1(2)$ transition with clearly resolved hyperfine sub-components.

Overall, 2603 lines have been assigned to the $\nu_4$ band with a standard deviation of ~0.00034 cm$^{-1}$ or 10 MHz, and 831 lines have been assigned to the $\nu_3+\nu_4-\nu_3$ hot band with standard deviation of ~0.00084 cm$^{-1}$ or 25 MHz. Table 1 compares the number of assigned lines in the two bands, and their corresponding standard deviation, from our comb-based FTS measurement to those based on previous FT-IR work [19, 20].

Table 1 Current status of the line positions of the $\nu_4$ band and the $\nu_3+\nu_4-\nu_3$ hot band of $CH_3I$ in the C-H stretch region.

| Technique | | | $\nu_4$ | | $\nu_3+\nu_4-\nu_3$ | |
|---|---|---|---|---|---|---|
| | Point Spacing | Ref. | Assignments | St. dev./ cm$^{-1}$ | Assignments | St.dev./ cm$^{-1}$ |
| FT-IR | 150 MHz | [19] | 500 | 0.005 | -- | -- |
| FT-IR | 162 MHz | [20] | 1850 | 0.00083 | 380 | 0.0013 |
| Comb-FTS | 11 MHz | This work | 2603 | 0.00034 | 831 | 0.00084 |

Tables 2 and 3 summarize the parameters of the $\nu_4$ band and the $\nu_3+\nu_4-\nu_3$ hot band obtained from our comb-based FTS measurements at 11 MHz sample point spacing and that of Anttila et al. [20] using standard FT-IR with 162 MHz resolution. Figure 6 shows the residuals of the least squares fit to the 2603 transitions assigned in our work as a function of $J$ and $K$ quantum numbers using the band parameters from our study [(panel (a)] and from the earlier work of Anttila et al. [43] [panel (b)]. Both simulations used the same set of ground state constants from Refs [40, 41]. As shown in this figure, the simulation based on our parameters leads to residuals randomly scattered around zero for transitions up to $J$=75 and $K$=12 quantum numbers, while that based on parameters from Anttila et al. shows larger systematic disagreements for higher $J$ numbers (note the difference in the y-axis scale range in the two panels). Our simulations show a much smaller deviation for the levels involving higher $J$ values, and hence the reported band parameters from our work are more accurate than from the earlier FT-IR measurements. The large standard deviation for transitions with higher $J$ numbers in Anttila et al. is attributed to the small difference in the reported rotational ($A$, $B$) and the centrifugal distortion ($D_J$, $D_{JK}$, $D_K$) constants. Such high fit sensitivity to small differences in band parameters is enabled by the high-resolution and precision capabilities of the comb-based Fourier transform spectroscopy.

Table 2 Parameters of the $\nu_4$ band of $CH_3I$ obtained from this study compared to that of Anttila et al. [20].

| Constants [cm$^{-1}$] | This Work | Anttila et al.[20] |
|---|---|---|
| $\nu_0$ | 3060.07846(3) | 3060.07890(6) |
| $A$ | 5.144014(2) | 5.144038(4) |
| $B$ | 0.25033919(3) | 0.25034228(7) |
| $D_J \times 10^7$ | 2.09373(9) | 2.09410(14) |
| $D_{JK} \times 10^6$ | 3.3039(5) | 3.3139(17) |
| $D_k \times 10^5$ | 9.238(2) | 9.271(7) |
| $\zeta$ | 0.066352(7) | 0.06655(10) |
| $\eta_J \times 10^7$ | -5.70(3) | -5.28(8) |
| $\eta_k \times 10^5$ | 4.33(2) | 4.43(7) |
| $q_+ \times 10^5$ | -1.0102(4) | -1.052(5) |
| Number of lines | 2063 | 1850 |
| Std. Dev | 0.00035 | 0.00083 |

Table 3 Parameters of the $\nu_3+\nu_4-\nu_3$ hot band of $CH_3I$ obtained from this study compared to that of Anttila et al. [20].

| Constants [cm$^{-1}$] | This Work | Anttila et al.[20] |
|---|---|---|
| $\nu_0$ | 3062.04686(5) | 3062.0474(2) |
| $A$ | 5.147074(9) | 5.14704(4) |
| $B$ | 0.24852518(2) | 0.24852(17) |
| $D_J \times 10^7$ | 2.1015(6) | 2.09410(14) |
| $D_{JK} \times 10^6$ | 2.886(6) | 3.0933(12) |
| $D_k \times 10^5$ | 7.643(3) | 7.830(5)e-5 |
| $\zeta$ | 0.068644(4) | .06865(6) |
| $\eta_J \times 10^6$ | -1.25(3) | -0.528(8) |
| $\eta_k \times 10^5$ | 2.85(2) | 4.43(7) |
| $q_+ \times 10^6$ | -9.12(2) | -9.44(5) |
| Number of lines | 831 | 380 |
| Std. Dev | 0.00084 | 0.0013 |

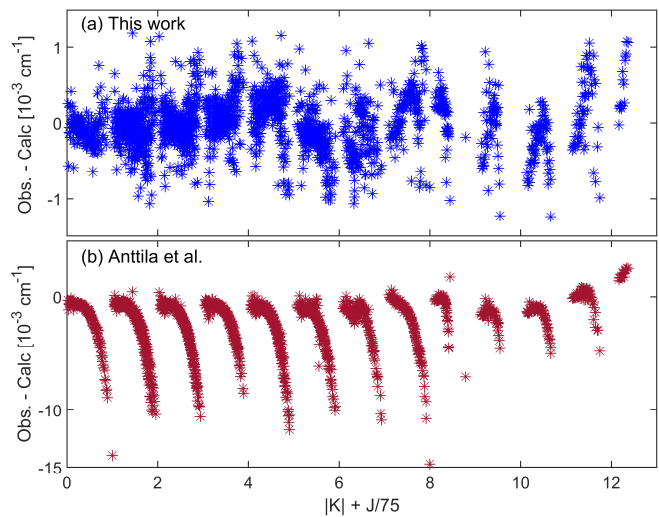

Fig. 6 Fit residuals (observed - calculated) as a function of the upper state K and J quantum numbers of the $\nu_4$ band for (a) our work and (b) using the band parameters of Anttila et al. [20].

### 3.3. Line intensities

The intensities of individual assigned lines of the $\nu_4$ band were found by applying a multispectral fit [44] to selected lines in the spectra measured at four different partial pressures of $CH_3I$, namely 0.11 mbar, 0.15 mbar, 0.20 mbar and 0.25 mbar, diluted with dry air to a total pressure of 10 mbar. To select these transitions, we first fit Voigt functions to the absorption lines in the spectrum measured at 0.11 mbar of pure $CH_3I$, same as used for the simulations above, with center frequencies, Lorentzian widths and intensities as free parameters. The fits were applied to the absorption lines separated from the nearest line by more than the Doppler full-width and whose peak absorption coefficient exceeded a threshold value of $1 \times 10^{-5}$ cm$^{-1}$, excluding those located within the congested Q sub-branches. Second, the retrieved line centers were compared to the line list obtained from the PGOPHER simulations. The closest simulated line was assigned to each fit line, if their positions agreed to within 30 MHz, in agreement with the residual shown in Fig. 6(a). Degenerate transitions, which correspond to the A1 and A2 symmetry of each K components of the K=3n (n=1,2,3,...) levels in the simulation, were treated as one line in the subsequent fitting. Simulated lines that were too weak to be observed and could obstruct correct matching of the simulated and observed absorption lines were filtered out prior to the assignment process. At the end of this process, a total of 838 lines were selected and assigned. Of these, 558 lines separated by at least 10 times the Doppler width (i.e. 960 MHz) were selected for the multispectral fitting.

For the multispectral fit we used a home-written non-linear fitting routine based on the Levenberg-Marquardt algorithm. Each line was modelled as a Voigt line shape with the linestrength and center frequency as fitting parameters. An example of a fit to the $^rR9(25)$ line is shown in Fig. 7(a). The line positions at 10 mbar were found to be shifted on average by about 2 MHz relative to the measurement of pure $CH_3I$ at 0.11 mbar. This could be accounted for by the pressure shift, but it is close to the uncertainty of 1.5 MHz on the line positions originating from the sub-nominal sampling procedure. The Lorentzian width was fixed to a global value that simultaneously maximized the fit quality on a subset of selected strong lines, which was done for two reasons. First, when the Lorentzian width was fit, its variation across the entire spectrum was random with standard deviation of 1%. Second, we observed a non-negligible instrumental broadening, possibly caused by phase noise of the DFG comb originating from noise on the current driver of the pump diode of the Yb-doper fiber amplifier. Previous measurements on samples of methane, as a test molecule, in the Doppler limit and at known concentrations indicated that the broadening can be modelled by fitting the Lorentzian component of the Voigt profile, and the resulting errors in the fitted line areas were lower than 5%. Since the Lorentzian component of the fit contains an unknown contribution of instrumental origin, we decided not to report the Lorentzian width value. A pressure offset of 19 μbar, attributed to an offset of the pressure sensor, was subtracted from all pressure readings as this improved the fit quality. Some of the strongest lines showed signs of saturation in the measurement at 0.25 mbar, which was therefore excluded from the fit for these lines.

We rejected the non-convergent fits, lines with signal to noise ratio lower than 25 at 0.11 mbar of $CH_3I$ in 10 mbar of air (which corresponds to linestrength lower than $0.4 \times 10^{-22}$ cm$^{-1}$/molec/cm$^2$) and fits with quality factor less than 30. The remaining 325 lines were visually inspected to further remove fits that were compromised by interference with weak absorption lines of the $\nu_4$ or the $\nu_3+\nu_4-\nu_3$ band, and/or by baseline issues due to artefacts not corrected by the baseline removal. We also discarded a dozen lines that appeared anomalously broad, perhaps as a result of unresolved hyperfine splitting. This finally left 157 lines, of which 50 are degenerate transitions with A1 and A2 symmetry. The intensities of those lines are shown in Fig. 7(b). The error bar in the figure shows the precision of the fit, which is on the 0.5% level. We estimate the total uncertainty on the line intensities to be 7%, where 5% comes from the instrumental broadening, 4% comes from the resolution of the pressure sensor, 1.3% comes from the uncertainty on the path length in the multipass cell, and 1% from the purity of the $CH_3I$ sample (all summed in quadrature with the fit uncertainty).

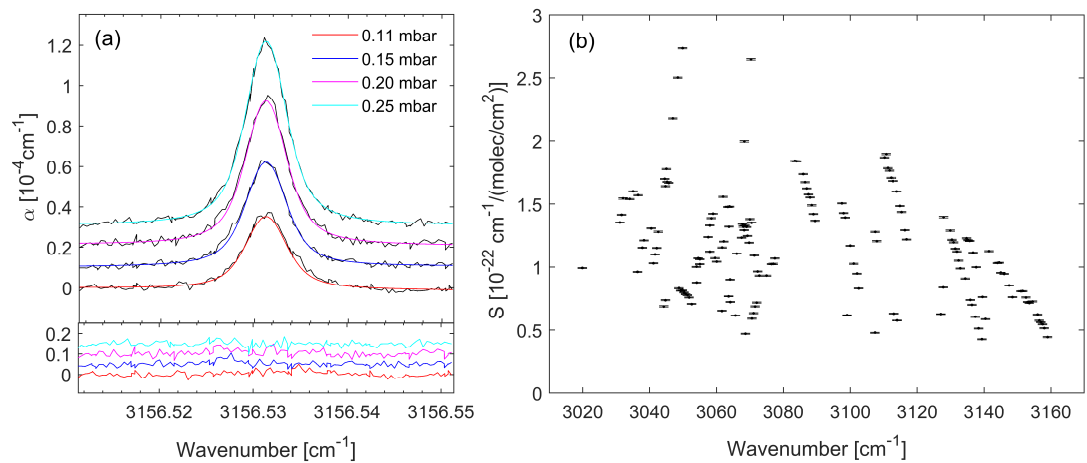

Fig. 7 (a) The multispectral fit to the degenerate $^rR9(25)$ line, showing the data from the measurements at four different partial pressures in black (offset incrementally upwards for clarity). The corresponding fits are shown in red, blue, purple and turquoise. The legend shows the $CH_3I$ partial pressures. The fit residuals are shown in the lower panel, offset vertically. (b) The fitted linestrengths for 157 well-resolved lines. The error bars display the fit precision.

## 4. Conclusions

Optical frequency comb Fourier transform spectroscopy is introduced as a platform for precision measurements of the lacking high-resolution spectra and the line shape parameters of trace halocarbons such as $CH_3I$. The ro-vibrational spectrum of $CH_3I$ in the entire region from 2800 to 3160 cm$^{-1}$ was measured simultaneously, covering three main features of the parallel vibrational overtone and combination bands located around 2850 cm$^{-1}$ and the two fundamental bands: the $\nu_1$ band, centered at 2971 cm$^{-1}$, and the $\nu_4$ band centered at 3060 cm$^{-1}$. Using the measured ro-vibrational spectrum in the range from the 3010 to 3160 cm$^{-1}$ together with the available microwave data [40, 41], we simulated the spectra of the $\nu_4$ band and the nearby weak $\nu_3+\nu_4-\nu_3$ hot band using PGOPHER. New assignment of these bands was introduced and compared to the earlier assignment of Anttila et al. based on standard FT-IR measurements with a resolution of 162 MHz [20]. A least square fit of the assigned transitions to the measured spectrum provided accurate upper state rotational constants of both bands. Overall, we assigned 2603 transitions to the $\nu_4$ band with st. dev. of 0.00034 cm$^{-1}$, and 831 transitions to $\nu_3+\nu_4-\nu_3$ hot band with st. dev. of 0.00084 cm$^{-1}$. The reported upper state band parameters are more accurate than those from the earlier FT-IR-based analysis [20], where 1830 transition with st. dev. 0.00085 cm$^{-1}$ were assigned to the $\nu_4$ band, and 380 transitions with st. dev. of 0.0013 cm$^{-1}$ were assigned to the $\nu_3+\nu_4-\nu_3$ hot band. The hyperfine splittings due to the $^{127}I$ nuclear quadrupole moment are clearly observable in the $\nu_4$ band, for the first time, for transitions with J $\leq 2 \times$ K, in agreement with recent analysis of the $\nu_6$ band [14]. We also reported intensities of 157 isolated lines in the $\nu_4$ band for the first time, using the Voigt line shape as a model in multispectral fitting. The new line list will serve as a future reference for further investigation of the spectral line parameters for public spectral databases such as HITRAN [45] and GEISA [46] as well as a basis for line selection in future monitoring applications of $CH_3I$.

## Acknowledgements

The authors thank Grzegorz Soboń for providing the polarization maintaining highly nonlinear fiber for the DFG source, Colin Western for providing help with simulating the spectrum in PGOPHER, and Isak Silander for help with setting up the vacuum system and gas supply. This project is financed by the Knut and Alice Wallenberg Foundation (KAW 2015.0159) and the Swedish Research Council (2016-03593).